\documentstyle[seceq]{ptptex}

\title{
 \begin{flushright}
  KOBE-TH-00-09\\
\end{flushright}
\vspace*{0.5cm} 
Implications of sigma models in the standard model and beyond
\footnote{Invited talk at the International Workshop on ``Possible existence 
of the $\sigma$-meson  and its implications to hadron physics", 
June 12-14, 2000, Yukawa Institute for Theoretical Physics, Kyoto, Japan.}
}

\author{
C.S. {\sc Lim}
}

\inst{
Physics Department, Kobe University, Nada, Kobe 657-8501, Japan
}

\abst{
After a brief introduction to the sigma model in QCD, we discuss how the sigma 
model can be relevant in the Standard Model. It is shown to be useful in the analysis of weak processes, such as the study of $\Delta \mbox{I} = 1/2$ rule, and in the analysis of the Higgs sector as well. The sigma model description is also shown to be quite useful in searching for the effects of New Physics via gauge boson self-interactions.}

\begin{document}

\maketitle

\section{Implications of sigma model in the Standard Model}

\subsection{Relevance of sigma model in QCD}

The sigma model was originally developed as an effective low energy theory of QCD. Suppose we have just two massless flavours of quarks, $(u, d)$ with 
$m_u = m_d = 0$. Then the QCD lagrangian has chiral symmetry $\mbox{SU(2)}_L \times \mbox{SU(2)}_R$, under independent unitary 
transformations of left- and right-handed quarks by the elements of SU(2), $U_L$ and $U_R$, respectively, 
\begin{equation} 
\pmatrix{ u \cr
          d}_{L} \ 
\rightarrow \ 
\pmatrix{ u' \cr
          d'}_{L} = U_L \ \pmatrix{ u \cr
            d}_{L}, \ \ 
\pmatrix{ u \cr
          d}_{R} \ 
\rightarrow \ 
\pmatrix{ u' \cr
          d'}_{R} = U_R \ \pmatrix{ u \cr
            d}_{R} . 
\end{equation}
Actually, such global symmetry is spontaneously broken by the presence of 
the vacuum expectation values (VEV), $<\overline{u}u> = <\overline{d}d> = O(\Lambda_{\mbox{QCD}}^{3})$, where $\Lambda_{\mbox{QCD}} \sim 200 \mbox{MeV}$ is a unique mass scale in QCD and is comparable to the pion decay constant, $f_{\pi} \sim 93 \mbox{MeV}$, roughly speaking. Though the ``vacuum condensations" break chiral symmetry, there remains a vectorial (parity conserving) $\mbox{SU(2)}_V$ symmetry with $U_L = U_R$, as the condensations are symmetric under the rotation between $u$ and $d$. Such symmetry is often called isospin symmetry. It is a general wisdom that when some symmetry is spontaneously broken there should appear some massless particles, called Nambu-Goldstone (N-G) bosons. The point which makes this statement so powerful is that 
the appearance of such massless particles is guaranteed just by the symmetry 
argument. Thus even though we know little about the non-perturbative aspects of QCD, such as the formation of bound states, the appearance of 
particles is of no doubt. Then a natural guess is that the N-G bosons should be identified with light hadrons with odd parity, i.e. pions. In reality, pions are not exactly massless as $u$ and $d$ quarks have tiny masses. The pions are iso-triplet pseudo-scalar bound states, 
\begin{equation}
\pmatrix{\overline{u} & \overline{d}}_{L} i \sigma^{a} \pmatrix{ u \cr
            d}_{R} + h.c.  \sim  \pi^{a} \ \ \ (a = 1 -3). 
\end{equation} 
We should have a chiral-partner of the pions, $\sigma \sim \overline{u}u + \overline{d}d$, whose confirmation is 
the main topics of this workshop, The sigma meson behaves as iso-singlet. 
Combining these 4 fields we get a matrix representation, $\Sigma = \sigma \mbox{I} + i \pi^{a} \sigma^{a}$, which transforms under the chiral transformation as $\Sigma \rightarrow \Sigma' = U_{L} \Sigma U_{R}^{\dagger}$. The low-energy phenomena are now described by the linear sigma model in terms of  this $\Sigma$ field, ${\cal L} = (1/4) {\mbox Tr} (\partial_{\mu} \Sigma^{\dagger})(\partial^{\mu} \Sigma) - V(\Sigma)$, where the potential term, denoted by $V$, should be invariant under chiral symmetry , while the kinetic term is easily seen to be chiral symmetric.
The VEV of sigma field, $< \Sigma > = < \sigma > \mbox{I}$, leaves $\mbox{SU(2)}_{V}$ as an unbroken symmetry. As only the sigma field is not N-G boson, it may be  treated as heavy at very low energies, and we get a theory without the field, that is non-linear sigma model.

\subsection{Relevance of sigma model in the Standard Model}

How can the sigma model be relevant in the Standard Model of particle physics ?  We may naively think that the typical mass scale in the Standard Model, $M_W$,  is much larger than $\Lambda_{\mbox{QCD}}$ or $f_{\pi}$, and the $\sigma$ model to describe the hadrons as bound states is no longer relevant in weak interacion processes. We, however, find that the  sigma model is still relevant even in the Standard Model. The reason is twofold:  

\noindent a. To describe low-energy weak processes of hadrons 
 
Even though in weak interaction the elementary processes of quarks and leptons 
are handled by the physics at short distances, such as $O(1/M_W)$ or $O(1/m_t)$, low energy weak processes of hadrons still need the sigma model in order to account for the confining QCD forces at longer distances. 

If we are interested in weak processes of hadrons made of 3 light flavours, $u, d, s$, the $\Sigma$ field is generalized to a $3 \times 3$ matrix, and in order to incorporate the weak-electromagnetic interactions the ordinary derivative 
has to be replaced by gauge-covariant derivative, $D_{\mu} \Sigma$. 
The kinetic term now reads as $(1/4) {\mbox Tr} (D_{\mu} \Sigma^{\dagger})(D^{\mu} \Sigma)$: gauged sigma model. Actually in the covariant derivative, weak gauge bosons, $W^{\mu}, Z^{\mu}$ 
are not suitable to appear, as they are so heavy that they have been decoupled from such low energy processes where the sigma model description is relevant. 
So the weak gauge bosons, together with heavier quarks like t quark, should have been integrated out (in the same philosophy as the ``Wilsonian renormalization"). The prescription goes as follows. First we calculate Feynman diagrams where heavy particles appear in the internal loops, to  get effective operators for light quarks. For instance famous penguin diagram provides an $\Delta \mbox{I} = 1/2$ four Fermi operator, 
$c_P \ (\overline{s}_L \gamma^{\mu} d_L ) \ (\overline{u} \gamma_{\mu} u 
+ \overline{d} \gamma_{\mu} d)$, where the Wilson coefficient $c_P$ is the function of heavy particle masses, the result of loop integration, and  contains the physics at short distances. Next we need to incorporate the physics at longer distances, namely we have to replace the effective operator by a corresponding one with respect to the $\Sigma$ 
field. We may achieve this by replacing the quark currents by the corresponding  currents of hadrons, which have the same transformation properties as the original quark currents under the chiral symmetry $\mbox{SU(3)}_L \times \mbox{SU(3)}_R $. For instance, the flavour changing neutral current, appearing in the penguine operator, $\overline{s}_L \gamma^{\mu} d_L$, can be replaced by 
\begin{equation}
\frac{1}{4} {\mbox Tr} \{ (i \partial_{\mu} \Sigma^{\dagger}) 
\pmatrix{0 & 0 & 0 \cr 
         0 & 0 & 0 \cr 
         0 & 1 & 0 } 
\Sigma \} . 
\end{equation}
In this way we get an effective interaction lagrangian with respect to hadrons.  As is well-known the physics of kaons, especially its processes with CP violation,  played an essential role for the foundation of the Standard Model. 
In kaon physics the $\Delta {\mbox I} = 1/2$ rule remains as a mystery. E.P. Shabalin and also T. Morozumi, A.I. Sanda and myself have made an observation that in the decay process, $K \rightarrow \pi \pi$, the sigma meson exchange should be important, though usually such processes are treated by use of non-linear sigma model \cite{rf:MOROZUMI}. We should note that the typical energy scale of the process, $m_K$, is not far from the expected sigma meson mass, or even comparable to, as suggested in this workshop. For more details of the work, see Ref.\cite{rf:AIS}.

\noindent b. Higgs as a sigma meson 
 
Another application of the sigma model in the Standard Model comes from the fact that the particle, which is responsible for the spontaneous breakdown of gauge symmetry, Higgs scalar, behaves just as the sigma meson, though it is no longer a bound state. The crucial point is that the spontaneous symmetry breaking, 
$\mbox{SU(2)}_L \times \mbox{U(1)}_Y \rightarrow \mbox{U(1)}_{\mbox{e.m.}}$ may be understood (in the Higgs sector of the theory) as the breaking of a chiral symmetry. Assuming isospin symmetry, $m_u = m_d$, the Yukawa coupling of Higgs doublet $\phi$ can be re-written in a matrix form, 
\begin{equation}
{\cal L}_{Y} = f \{ \pmatrix{\overline{u} & \overline{d} }_{L} \phi u_{R} 
                  +  \pmatrix{\overline{u} & \overline{d} }_{L} \tilde{\phi} d_{R} \} 
               + \ h.c. 
             = f \ \pmatrix{\overline{u} & \overline{d} }_{L} \ 
                   \pmatrix{\phi & \tilde{\phi}} \pmatrix{u \cr 
                                                          d}_{R} 
               + \ h.c.,  
\end{equation}
where $f$ is the common Yukawa coupling and $\tilde{\phi} \equiv i \ \sigma_{2} \phi^{\ast}$. This argument suggests that a $2 \times 2$ matrix 
\begin{equation}
\Sigma \equiv \pmatrix{\phi & \tilde{\phi}} = 
 \pmatrix{\phi^{0} & -\phi^{+}  \cr 
          \phi^{-} &  \phi^{0\ast}}
= \frac{1}{\sqrt{2}} (h \ \mbox{I} + i \sigma^{a} G^{a}), 
\end{equation}
just behaves as the $\Sigma$ field in the effective theory of QCD with 2 flavours. Here 
$h$ denotes the Higgs field and $G^{a} \ (a = 1,2,3)$ denote 
massless ``would-be" N-G bosons. The $\Sigma$ has exactly the same transformation property as in QCD under chiral transformation $\mbox{SU(2)}_L \times  \mbox{SU(2)}_R$, $\Sigma \rightarrow \Sigma' = U_{L} \Sigma U_{R}^{\dagger}$. Now the Higgs sector of the theory can be written in terms of a (gauged) linear sigma model, ${\cal L}_{H} = (1/2) \mbox{Tr} (D_{\mu} \Sigma^{\dagger} 
D^{\mu} \Sigma) - (1/2) \mu^{2} \mbox{Tr} (\Sigma^{\dagger} \Sigma) 
- (1/4) \lambda \{ \mbox{Tr} (\Sigma^{\dagger} \Sigma) \}^{2}$. The VEV $ <\Sigma> = \frac{v}{\sqrt{2}} \ \mbox{I}$ causes spontaneous breaking, $\mbox{SU(2)}_L \times  \mbox{SU(2)}_R \rightarrow \mbox{SU(2)}_{V}$, leaving a vector-like symmetry $\mbox{SU(2)}_{V}$, which is often called custodial symmetry. An important consequence of this symmetry is concerning gauge boson masses. The gauge bosons $W^{a}_{\mu}$ behave as triplet under the both of $\mbox{SU(2)}_{L}$ and the custodial symmetry $\mbox{SU(2)}_{V}$. Thus we get degenerate gauge boson masses, $M_{W^1} = M_{W^2} = M_{W^3}$, and 
equivalently the unity of so-called $\rho$-parameter, $\rho \equiv 
\frac{M_{W}^{2}}{M_{Z}^{2} \mbox{cos}^{2} \theta_{W}} = 1$, at the classical level. In contrast to the case of QCD, where the sigma particle has a dynamical mass of the order of $\Lambda_{QCD}$, the Higgs mass is a free parameter and if the mass is much larger than the energy scale we consider, the description in terms of non-linear sigma model will become relevant.

\section{The implication of sigma model in the physics beyond the Standard Model}

If the Standard Model is regarded to be valid up to very high energy scale, $\Lambda$, the model suffers from a serious problem. Namely, 
the Higgs scalar mass-squared gets a quantum correction of $O(\Lambda^{2})$, 
and a fine tuning of the bare mass is needed: so-called hierarchy problem. 
In the trial to solve the problem, models of physics beyond the Standard 
Model, ``New Physics", have been devised. The most popular is supersymmetric theories (Minimal Supersymmetric Standard Model, Superstring Theory). There is also an idea that the Higgs field is actually a bound state of hypothetical fermions.  The typical example is Technicolor theory. This is a scaled up version of QCD, where $\Lambda_{\mbox{QCD}}, \ f_{\pi}$ are scaled up to $\Lambda_{\mbox{TC}} \sim 1 \mbox{TeV}, \ v \sim 200 \mbox{GeV}$ \ ($v$ is the VEV in the Standard Model). Just as in QCD the Higgs sector is described by a sigma model. But as the Higgs mass is expected to be rather heavy, $m_{h} \sim 1 \mbox{TeV}$, the theory may be well described by a gauged non-linear sigma model, ${\cal L}_{h} = \frac{v^{2}}{4} \mbox{Tr} (D_{\mu} U^{\dagger} D^{\mu} U) \ \ (U = \mbox{exp} (iG^{a} \sigma^{a}/v))$, at $ E \ll \Lambda_{\mbox{TC}}$. 
Recently there has appeared another interesting possibility that the hierarchy problem may be naturally solved if our space-time is not 4-dimensional but has some extra-dimensions \cite{rf:EXTRA}. 

\subsection{Oblique parameters and New Physics} 

The models of New Physics have their own new heavy particles, such as super-partners, techni-hadrons, non-zero Kaluza-Klein modes, etc.. 
It often happens that these heavy particles do not directly couple with quarks and leptons, though they affect electro-weak processes of quarks and leptons indirectly through the quantum correction to the gauge boson self-energies. Such quantum correction is called oblique correction. The oblique correction due to new heavy particles is known to be described by just 3 parameters, $S, T$ and $U$ 
\cite{rf:PESKIN}:
\begin{eqnarray}
\alpha S & = & -4e^{2} \pi_{3Y}'(0), \nonumber \\ 
\alpha T & = & \Delta \rho = \rho - 1 = \frac{e^{2}}{M_{W}^{2} \mbox{sin}^{2}
\theta_{W}} \{\pi_{11}(0) - \pi_{33}(0)\}, \\ 
\alpha U & = & 4e^{2} \{\pi_{11}'(0) - \pi_{33}'(0)\},  \nonumber 
\end{eqnarray}
where $\alpha = e^{2}/(4\pi )$, $\pi '(q^{2}) \equiv \frac{d\pi (q^{2})}{d q^{2}}$ and $\pi_{ij}(q^{2})$ is a Lorentz scalar part of the vacuum polarization tensor: $\pi_{ij}^{\mu \nu}(q^{2}) = g^{\mu \nu} \pi_{ij}(q^{2}) + q^{\mu}q^{\nu} \mbox{term} \ \ (i,j = 1,2,3)$. 

\subsection{Oblique parameters and chiral symmetry breaking} 

These oblique parameters, $S, T$ and $U$, turn out to be closely related with 
the property of the chiral symmetry breaking \cite{rf:INAMI}, and therefore with the sigma model. 
We first note that these parameters behave under the chiral symmetry $ \mbox{SU(2)}_L \times  \mbox{SU(2)}_R$ as the following representations: \  $S$: (3,3), \  $T, U$ :  (5,1). Let us note that the $S$-parameter is concerned with the mixing in the kinetic term of $W_{\mu}^{3}$ and $W_{\mu}^{Y}$ gauge bosons of $\mbox{SU(2)}_{L}$ and $\mbox{U(1)}_{Y}$, i.e. $\pi_{3Y}'(0)$.  The weak-hypercharge can be 
decomposed as $Y = I_{3R} + \frac{B-L}{2}$ ($I_{3R}$ is generated by $\mbox{SU(2)}_R $), and we have taken only the part of $I_{3R}$, for brevity. 
By taking the tensor products of repr.s of each chiral symmetry we get the repr. of the custodial symmetry: \ $S$ : 1 + 5, \ \ $T, U$ : 5. 
Thus we find the following facts. (a) \ If the chiral symmetry $\mbox{SU(2)}_L \times  \mbox{SU(2)}_R$ is exact, all parameters vanish, i.e. $S = T = U = 0 $. \ (b) \ If the chiral symmetry is broken into the custodial symmetry, $\mbox{SU(2)}_L \times  \mbox{SU(2)}_R \rightarrow \mbox{SU(2)}_{V}$, $S \neq 0$, while 
$T = U = 0$. The contribution of techni-hadrons drops into this category and it is a famous story that the original version of Technicolor theory has been ruled out by the stringent experimental constraint on the $S$-parameter \cite{rf:PESKIN}. \ (c) \ If even the custodial symmetry is broken (e.g. by $m_{t'} \neq m_{b'}$, with $t', b'$ being possible (?) 4-th generation quarks), $S, T, U \neq 0$. 

\subsection{Sigma model description} 

Though the gauge symmetry $\mbox{SU(2)}_{L} \times \mbox{U(1)}_{Y}$ is 
spontaneously broken, every quantum correction to the observable, which was absent in the original theory (oblique corrections, the amplitudes of Flavour Changing Neutral Current, etc.), should be described by an effective gauge invariant operator, once we include Higgs doublet, even if what we are interested in is 
not concerning the Higgs itself, like in oblique corrections. Such operator is inevitably higher dimensional ($d > 4$) ``irrelevant" operator, and the sum of these operators will form a gauged sigma model. 
For instance, the $S$-parameter is described by an irrelevant ($d = 6$) operator, 
$c_{s} \phi^{\dagger} (W_{\mu\nu}^{a} \sigma^{a}) \phi \cdot B^{\mu\nu}$, 
where $W_{\mu\nu}^{a}$ and $B_{\mu\nu}$ are field strengths of 
$\mbox{SU(2)}_{L}$ and $\mbox{U(1)}_{Y}$ gauge bosons. After the spontaneous breakdown the Higgs doublet $\phi$ can be replaced by the VEV $v$. It is now clear that the breaking of chiral symmetry by the $v$ leads to the $S$-parameter, i.e. $S \sim c_{s} v^{2}$. 

The bonus of using such sigma model description is that we can immediately realize that the New Physics contribution appears not only in gauge boson 2-point functions, like oblique corections, but also in triple gauge boson vertices (TGV). For instance in the operator in charge of the $S$-parameter, $W_{\mu\nu}^{a} $ contains a term quadratic in the gauge bosons (non-Abelian nature), leading to a term $W^{+}_{\mu}W^{-}_{\nu}B^{\mu\nu}$. A detailed analysis \cite{rf:TAKEUCHI} shows that the New Physics contribution to TGV is described by just 4 parameters, independent of $S, T, U$-parameters. 

\subsection{Decoupling or non-decoupling} 

To be more precise, there are two types of the New Physics contributions. 
The first case is that the contributions of new heavy particles (with typical mass $M$) are suppressed by the inverse powers of $M$: \ the case of decoupling. 
The typical examples are the contributions of super-partners, GUT particles. The second case is that the contributions of new heavy particles are not suppressed by their masses: \ the case of non-decoupling. The typical examples are the contributions of 4-th generation quarks and leptons, techini-hadrons.

In the case of decoupling, the gauged sigma model is well approximated solely by the $d = 6$ operators. This is simply because the Wilson coefficient of an 
 operator behaves as $M^{4-d}$ ($d$ being the mass dimension of the operator in question), and the coefficients of operators with $d > 6$ are relatively suppressed by the powers of $v^{2}/M^{2}$, compared with the coefficients of $d=6$ operators. 

In the case of non-decoupling, however, such suppression factor of $v^{2}/M^{2}$  does not exists. This is because in the non-decoupling case, in contrast to the case of decoupling mentioned above, heavy particles get their masses not from some new large mass scale $M$, but from the spontaneous breaking of gauge symmetry $v$. Hence, in the case of non-decoupling, higher dimensional gauge invariant  operators, such as $ \phi^{\dagger} (W_{\mu\nu}^{a} \sigma^{a}) \phi \cdot B^{\mu\nu} \cdot (\phi^{\dagger}\phi )^{n} \ (n = 1,2,..)$ all contribute to the same observable. Utilizing non-linear representation of the Higgs doublet, 
$\phi \rightarrow U \cdot v \ (U = \mbox{exp}(iG^{a}\sigma^{a}/v))$, turns out to be 
useful in order to avoid the cumbersome problem caused by the presence of such higher dimensional operators. (Let us note that in this repr. $\phi^{\dagger} \phi = v^{2}$ is just a c-number.) This non-linear sigma model description is technically useful, even if Higgs is not so heavy. 
For instance the $T$-parameter or $\Delta \rho = \rho - 1$ is described by the Wilson coefficient of the operator $\mbox{Tr} \{ (U^{\dagger} iD^{\mu} U) \sigma_{3}  (U^{\dagger} iD_{\mu} U) \sigma_{3} \}$. Appelquist and Wu \cite{rf:APPELQUIST} have shown that New Physics contributions to gauge boson 2- and 3-point functions are described by 7 operators of non-linear sigma model, which is consistent with our result \cite{rf:TAKEUCHI}. Recently, we have succeeded to generalize this kind of argument and show, by making use of the idea of Wilsonian renormalization to get the effective action, that all New Physics contributions to gauge boson self-interactions are concentrated in the gauge boson 2-, 3-, and 4-point functions, but not any more \cite{rf:LIM}.

\section*{Acknowledgements}
I would like to thank my collaborators, H.Hatanaka, T.Inami, T.Morozumi, A.I.Sanda, B.Taga, B.Takeuchi, M.Tanabashi and A.Yamada, for their fruitful collaborations, which my talk was based upon. This work has been supported by the Grant-in-Aid for Scientific Research (12640275) from the Ministry of Education, Science and 
Culture, Japan.


\begin{thebibliography}{99}
\bibitem{rf:MOROZUMI}
E.P.~Shabalin, 
        Sov.~J.~Nucl.~Phys. \ {\bf 48} (1988), 173. 
\\
T.~Morozumi, C.S.~Lim and A.I.~Sanda, 
        Phys.~Rev.~Lett. \ {\bf 65} (1990), 404.   
\bibitem{rf:AIS}
A.I.~Sanda, 
        in these proceedings.
\bibitem{rf:EXTRA}
N.~Arkani-Hamed, S.~Dimopoulos and G.~Dvali, 
        Phys.~Lett. \ {\bf B429} (1998), 263. 
\\
H.~Hatanaka, T.~Inami and C.S.~Lim, 
        Mod.~Phys.~Lett. \ {\bf A13} (1998), 2601.
\\
L.~Randall and R.~Sundrum, 
        Phys.~Rev.~Lett. \ {\bf 83} (1999), 3370.
\bibitem{rf:PESKIN}
M.~Peskin and T.~Takeuchi, 
        Phys.~Rev.~Lett. \ {\bf 65} (1990), 964.
\bibitem{rf:INAMI}
T.~Inami, C.S.~Lim and A.~Yamada, 
        Mod.~Phys.~Lett. \ {\bf A7} (1992), 2789. 
\bibitem{rf:TAKEUCHI}
T.~Inami, C.S.~Lim, B.~Takeuchi and M.~Tanabashi, 
        Phys.~Lett. \ {\bf B381} (1996), 458. 
\bibitem{rf:APPELQUIST}
T.~Appelquist and G.-H.~Wu, 
        Phys.~Rev. \ {\bf D48} (1993), 3235.
\bibitem{rf:LIM}
C.S.~Lim and B.~Taga, 
         Prog.~Theor.~Phys. \ {\bf 103} (2000), 795. 
\end{thebibliography}
\end{document}